\documentclass[aps,showpacs,nofootinbib]{revtex4-1}
\usepackage{slashed}
\newcommand{\bb}{\begin{eqnarray}}
\newcommand{\ee}{\end{eqnarray}}
\begin{document}

\title{Fermion regularization, fermion measure and axion fields}
\author{P. Mitra}
\affiliation{Saha Institute of Nuclear Physics,\\
1/AF Bidhannagar\\ Calcutta 700064} 
%\date{1504.07936}
%\widetext
\begin{abstract}
Axion fields were originally introduced to control CP violation
due to the $\theta$ term in QCD. Pauli-Villars
regularization, or the use of a {\it parity symmetric} 
fermion measure, however, preserves CP in the fermion sector. 
A CP violation arising from the $\theta$ term
can then be neutralized in a natural way by setting $\theta$ equal to zero.
\end{abstract}
%\pacs{12.38.Lg, 14.80.Va}
\maketitle
%\vfill
%\section{Introduction}
The discovery of the Higgs boson put a final stamp of confirmation on the
standard model of high energy physics, but the axion, which is also very
widely expected, has not yet been seen. It is not quite a part of the
standard model, but is popularly connected with CP violation.
While CP violation is observed in the weak interactions, it is not known to
occur in other processes. However, 
%complex mass terms for quarks involving phases containing $\gamma^5$ and 
the so-called $\theta$ term $\theta F^{\mu\nu}\tilde F_{\mu\nu}$
arising from instantons indicates
the possibility of CP violation in the strong
interactions. As this has not been observed,
modifications of QCD have been proposed to suppress the CP violation.
The Peccei-Quinn hypothesis [1] introduced an artificial chiral symmetry
in an attempt to remove CP violation in the strong interactions.
Unfortunately, it leads to the occurrence of a light
pseudoscalar particle, the axion [2], which has not been detected in
spite of elaborate searches [3]. This may simply be like the delay in the
detection of the Higgs particle, but it may not be an experimental
failure at all. It may be the case that this is not the right way of
explaining the absence of CP violation in the strong interactions.
An alternative explanation [4] which does not require axions has indeed been
developed. It turns out that a complex quark mass term does not violate CP.
We wish to go further here and demonstrate that the axion field [1] does 
not generate CP violation in the fermion sector and cannot cancel CP violation
caused by a $\theta$ term if a natural regularization or measure is chosen.

While chiral symmetry of the fermion action is broken by a quark mass, a new 
chiral symmetry can be manufactured by letting a new pseudoscalar field 
${\varphi}$ absorb the chiral transformation. The mass term is then replaced
by
\bb\bar\psi m e^{i{\varphi}\gamma_5}\psi,\ee
which is invariant under the transformation
\bb\psi\rightarrow e^{i\alpha\gamma_5}\psi,\quad
\bar\psi\rightarrow \bar\psi e^{i\alpha\gamma_5},
\quad\varphi\rightarrow{\varphi}-2\alpha,\ee
which is also a symetry of the kinetic terms of the action but is anomalous. 
There have been variations on this theme.
The original interaction introduced by Peccei and Quinn [1] was of the form
\bb
\bar\psi [\Phi{1+\gamma^5\over 2}
+ \Phi^\dagger{1-\gamma^5\over 2}]\psi,\ee
where $\Phi$ is a complex scalar field with a symmetry breaking potential.
The chiral symmetry transformation is 
\bb\Phi\to e^{-2i\alpha}\Phi,\quad
\psi\to e^{i\alpha\gamma_5}\psi,
\quad\bar\psi\to \bar\psi e^{i\alpha\gamma_5}.\ee
$\Phi$ may be taken to be of the form $\rho e^{i{\varphi}}$.
The amplitude $\rho$ of the scalar field acquires a vacuum expectation value
because of symmetry breaking, which provides a massive boson.
The phase ${\varphi}$ is the zero mode of the potential 
and provides the Goldstone boson.
This is the axion, which acquires
a mass because of the quark masses, but does not appear to exist.
How does this mechanism claim to remove P and T violation?
It is usually believed that the exponential factor containing the axion field
can be absorbed in the fermion fields by a local chiral transformation,
which then produces a term $\varphi F^{\mu\nu}\tilde F_{\mu\nu}$ because of
the anomaly. The vacuum expectation value of the axion field could then nullify
the $\theta$ in the term $\theta F^{\mu\nu}\tilde F_{\mu\nu}$,
thus removing the CP violation due to $\theta$.

That analysis was done before the r\^ole of
measures in anomalies came to be fully appreciated.
The anomaly appears in regularized theories, so we shall use an explicit
regularization before carrying out the problematic chiral transformation
involving the axion field.
Does the $\theta$ term-like structure show up?

A convenient gauge invariant way to regularize the theory is to introduce
Pauli-Villars fields.   The Lagrangian density
\bb
\bar\psi[i{\slashed D} - m]\psi,%e^{i{\varphi}\gamma_5},
\ee
where $D$ is the covariant derivative and involves the gluon fields as
we are dealing with strong interactions, goes over to
\bb
\bar\psi[i{\slashed D} - m]\psi+%e^{i{\varphi}\gamma_5},
\bar\chi[i{\slashed D} - M]\chi.%e^{i{\varphi}\gamma_5}.
\ee
Here, $\chi$ is a bosonic spinor field, whose mass $M$ is to be taken to 
infinity. In the presence of the axion, the coupling has to be introduced:
\bb
\bar\psi[i{\slashed D} - me^{i{\varphi}\gamma_5}]\psi+
\bar\chi[i{\slashed D} - Me^{i{\varphi}\gamma_5}]\chi.
\label{2}\ee
Let us now carry out a 
local chiral transformation by $e^{-i\varphi\gamma_5/2}$ on $\psi,\chi$. 
The Jacobian of the measure is trivial in the regularized theory [4].
The phase factors disappear, but the kinetic terms produce derivatives
of $\varphi$.  The effective action arising from fermion integration
depends only on the differentiated axion field apart from the gauge field.
Therefore in this regularization there is no possibility of a term
$\varphi F^{\mu\nu}\tilde F_{\mu\nu}$ in the effective action which could
violate CP and
nullify the CP violating effect of the $\theta$ term.  %\hfill QED

The result goes against the belief that the exponential factor containing
the axion field is equivalent to a $\varphi F^{\mu\nu}\tilde F_{\mu\nu}$
term. This is because an explicit regularization has been used here. This
regularization does not reproduce the effect produced by the popular
choice of the measure of fermion integration which involves only an implicit
regularization and leads to the above term.
Does this mean that one of these calculations is wrong? Both are right
as far as calculations go, but we shall argue that only one is
appropriate in the circumstances.

There is an underlying symmetry. We first note that the axion coupling
conserves parity when $\varphi$ is a pseudoscalar field. Taking 
symmetry breaking into account, we can still preserve parity if 
the shifted field $\varphi'\equiv\varphi-\varphi_0$ 
is now transformed like a pseudoscalar,
where we denote the vacuum expectation value of $\varphi$ by $\varphi_0$.
One way of seeing this is to absorb $\varphi_0$ in new $\gamma$ matrices
\bb
\tilde\gamma^\mu\equiv\gamma^\mu e^{i\varphi_0\gamma_5},
\ee
which satisfy all requirements on $\gamma$ matrices. Note that
products of two $\gamma$ matrices are unchanged in this redefinition and
\bb
\bar\psi[i{\slashed D} - m e^{i{\varphi}\gamma_5}]\psi=
\psi^\dagger[i \tilde\gamma^0\tilde\gamma^\mu D_\mu - m 
\tilde\gamma^0e^{i{\varphi'}\gamma_5}]\psi.
\ee
Instead of using new $\gamma$ matrices, one may equivalently check the
invariance of $\bar\psi m e^{i\varphi\gamma_5}\psi$ under a new parity
transformation
\bb
\psi(\vec x)&\to &\gamma^0 e^{i\varphi_0\gamma_5}\psi(-\vec x)\nonumber\\
\bar\psi(\vec x)&\to &\bar\psi(-\vec x)\gamma^0 e^{-i\varphi_0\gamma_5},
\ee
which is also consistent with the kinetic part of the action.

The fermion action at the classical level thus has a parity symmetry. Upon 
quantization, two things can happen. Either there is no regularization which 
preserves this parity symmetry, in which case one says that the parity is 
anomalous; or, there are regularizations which preserve the symmetry,
in which case they should be used if artefacts are to be avoided. 
The regularization (\ref{2}) clearly exhibits the above parity when
$\varphi'$ is transformed like a pseudoscalar and $\chi$ is transformed
in the same way as $\psi$ under parity. It is therefore the natural 
regularization.

Instead of an explicit regularization, there is an
alternative measure approach with an implicit regularization.
The usual fermion integration measure, in this approach, fails to
respect the above symmetry. However, there exist choices of the measure which
preserve the symmetry [5,6]. These measures are dependent on the
axion field. Although a chiral transformation tends to
produce a Jacobian factor involving the angle of the chiral rotation,
these measures change also when the dependence on the axion field
has to be adjusted and the factors connected with these two changes cancel out.
This was discussed for a chiral phase in the complex mass term of
the fermion, {\it i.e.,} a constant axion, in [5], and for an axion field
in [6]. Instead of expanding the fermion field [7] as
\bb
\psi=\sum_n a_n\phi_n, \bar\psi=\sum_n \bar a_n\phi_n^\dagger,
\ee
where $\phi_n$ are eigenfunctions of ${\slashed{D}}$, parity symmetry demands
\bb
\psi=e^{-\frac12 i\varphi\gamma_5}\sum_n a_n\phi_n, \bar\psi=\sum_n \bar a_n\phi_n^\dagger e^{-\frac12 i\varphi\gamma_5}.
\ee
The fermion measure $\prod_n da_n d\bar a_n$
is gauge invariant and also parity invariant even when symmetry
breaking occurs and $\varphi_0$ is nonvanishing.
The consequence of this measure involving the axion
field is that this field is not transferred from the fermion sector of the
action to the gauge field sector
by the chiral transformation, as there is an additional
change in the measure which compensates the gauge field sector.
The local chiral transformation
replacing $e^{i{\varphi}\gamma^5/2}\psi$ by $\psi$
essentially removes $\varphi$ from both the measure and the action,
except for one residue: the kinetic term
of the fermion field is invariant only under global chiral transformations
and not local ones, so that a derivative term
results: $\bar\psi\gamma^\mu\gamma^5(\partial_\mu{\varphi})\psi$.
The upshot is that a ${\theta}$ term in the gauge
field sector cannot be cancelled by a ${\varphi}$ term
when this parity symmetric measure is employed. Nonsymmetric measures are 
unnatural in a technical sense.

One may wonder how much freedom one has in choosing regularizations or
measures. Different regularizations have of course been used in the past
and it is known that all do not lead to the same result. A key point is that
symmetries of the action should be sought to be preserved by the
regularization. Thus one is always looking for Lorentz invariant, gauge
invariant regularizations for theories with actions having such symmetries.
Bypassing this requirement leads to artificial results. It turns out that the 
regularization can be made parity invariant. 
In the case of measures the requirement is not so well
known because explicit measures are not often used.
However, if a symmetry of the action cannot be preserved by any measure,
it means that there is an anomaly. When there is no anomaly, there is no
reason to go for a measure that needlessly violates a symmetry of the action.
Gauge noninvariant or Lorentz violating measures have never been used
in theories where the actions have such symmetries. We know that in
the case of parity too, an invariant measure exists. If the measure is not
constrained to respect the symmetries of the action, it is possible to violate
all symmetries of the action simply by choosing non-invariant
measures. In particular, the action with a real mass term can give rise
to CP violation through the misuse of a measure not invariant under parity.
That is clearly artificial. The moral is that the measure,
like the regularization, has
to be sought to have the symmetry of the action as far as possible.

There may be an apprehension that a measure is predetermined and not in
one's hands. If a field is already quantized, one indeed cannot go and alter
the measure, but if one is looking at an action, which is basically a classical
concept, a measure has to be chosen. In the case at hand, one must clearly
understand where one stands. The quark mass comes not from the fermion sector or
the gauge sector but from scalar fields. For completeness, one may start
with massless quarks -- coupled to scalar fields. The vacuum expectation value
of the scalar generates both the mass and the chiral phase $\varphi_0$.
Note that the {\it same} phase arises in the mass term and in interaction terms
of the quarks with the scalars. So the same parity symmetry exists for both.
The quark may be considered to be quantized after that. At this stage a
regularization has to be chosen and the Pauli-Villars regularization may
be chosen. Alternatively, if an explicit measure is chosen, it is to be chosen 
with care, {\it  -- to preserve its symmetrical shape} .

%%%%%%%%%%%%%%%%%%%%%%%%%%%%%%%%%%%%%%%%%%%%%%%%%%%%%%%%%%%%%%%%%%%%%%%
%\section{Conclusion}
\bigskip

To sum up, the axion scheme of Peccei and Quinn was
invented to engineer the suppression of strong CP violation.
We have pointed out a parity symmetry of the classical fermion action including
the axion. This symmetry is preserved by explicit regularization and also in the
measure approach with implicit regularization unless one chooses to break it.  
As long as one respects this symmetry,
the axion field cannot produce a CP violating $\theta$-like term to compensate 
the CP violation caused by a $\theta$.
Naturally, $\theta$ itself may be set equal to zero [4] to avoid CP violation.

What remains finally is a concern about the need for axions. Even if there is
no r\^{o}le for axion fields in controlling CP violation,
they may just happen to exist: this can presumably be determined by experiment
and is beyond the reach of field theory.
%%%%%%%%%%%%%%%%%%%%%%%%%%%%%%%%%%%%%%%%%%%%%%%%%%%%%%%%%%%%%%%%%
%%%%%%%%%%%%%%%%%%%%%%%%%%%%%%%%%%%%%%%%%%%%%%%%%%%%%%%%%%%%%%%%%%%%
\section*{REFERENCES}
\begin{enumerate}
%\vspace{-1.5cm}
\item R. Peccei, \& H. Quinn, Constraints imposed by CP conservation
in the presence of instantons, {\it Phys. Rev.} {\bf D16}, 1791 (1977).
\item S. Weinberg,  A new light boson?
{\it Phys. Rev. Letters} {\bf 40}, 223 (1978).
%\item H. Banerjee, D. Chatterjee \& P. Mitra,  Is there a strong
%CP problem? {\it preprint SINP/TNP/90-5 (1990)}; 
%incorporated into Sec. V of} Dirac fermion in euclidean metric and the U(1) and
%strong CP problems. 
%{\it Zeit. f\"{u}r Phys.} {\bf C62}, 511-520 (1994).
%\item P. Sikivie,  Axion Searches,
%{\it Nucl. Phys. Proc. Supp.} {\bf 87}, 41 (2000).
\item K. A. Olive et al., Review of Particle Physics,
{\it Chin. Phys. } {\bf C38}, 090001 (2014).
\item H. Banerjee, D. Chatterjee \& P. Mitra,
Is there still a strong CP problem? %hep-ph/0012284,
{\it Phys. Letters} {\bf B573}, 109 (2003)
%\item E. Elizalde, S. D. Odintsov, A. Romeo, A. A. Bytsenko and S. Zerbini,
%{\it Zeta regularization techniques with applications}
%(World Scientific, Singapore, 1994)
%\item M. Reuter, Chiral anomalies and zeta-function regularization,
%{\it Phys. Rev.} {\bf D31} (1985) 1374
%\item K. Wilson, Confinement of quarks, {\it Phys. Rev.} {\bf D10}, 2445 (1974)
%\item E. Seiler,  \& I. O. Stamatescu, Lattice fermions and $\theta$ vacua,
%{\it Phys. Rev.} {\bf D25}, 2177; {\bf D26}, 534E (1982).
\item P. Mitra,
Complex fermion mass term, regularization and CP violation, {\it J. Phys.} {\bf A40},  F525 (2007).
\item P. Mitra, Fermion measure and axion fields, arXiv:1504.07936v1
\item K. Huang, {\it Quarks, Leptons \& Gauge Fields}, (World Scientific,
Singapore, 1992)
%\item Lewis Carroll, The Hunting of the Snark (1876)
%\item P. Mitra,  {\it Symmetries and symmetry breaking in field theory}, 
%CRC Press, Florida (2014)
\end{enumerate}

\end{document}